\documentclass[11pt]{article}

\usepackage[latin1]{inputenc}
\usepackage{graphicx}
\usepackage{color}
\usepackage{graphicx}
\usepackage{comment}
\usepackage[mathscr]{euscript}
\usepackage{algorithm}
\usepackage[noend]{algpseudocode}
\usepackage{enumitem}
\usepackage{lscape}
\usepackage{makecell}
\usepackage{lscape}
\usepackage{verbatim}
\usepackage[colorlinks=true, linkcolor=blue, citecolor=blue, urlcolor=blue]{hyperref}
\usepackage{booktabs, multirow}
\usepackage{amssymb,amsmath}
\usepackage{float}
\usepackage{xcolor}
\usepackage[round]{natbib}
\usepackage{chngcntr}
\usepackage{authblk}

\def\0{\bf 0}
\def\1{\bf 1}

\def\I{\bf I}
\def\D{\bf D}

\def\RR{\bf R}
\def\W{\bf W}

\def\U{\bf U}

\def\X{\bf X}

\def\Z{\bf Z}

\def\pphi{\mbox{\boldmath{$\phi$}}}
\def\xxi{\mbox{\boldmath{$\xi$}}}
\def\ggamma{\mbox{\boldmath{$\gamma$}}}
\def\ddelta{\mbox{\boldmath{$\delta$}}}

\begin{document}


\title{\textbf{Suicide Mortality in Spain (2010-2022): Temporal Trends, Spatial Patterns, and Risk Factors}}

\author[1,2,3]{Adin, A.}
\author[1,2]{Retegui, G.}
\author[4,5,6]{S\'anchez Villegas, A.}
\author[1,2,3]{Ugarte, M.D.}

\affil[1]{\small Department of Statistics, Computer Sciences and Mathematics, Public University of Navarre, Spain.}
\affil[2]{\small Institute for Advanced Materials and Mathematics, InaMat$^2$, Public University of Navarre, Spain.}
\affil[3]{\small Department of Mathematics, Centro Asociado UNED-Pamplona, Spain.}
\affil[4]{\small Institute for Innovation and Sustained Development in Food Chain (ISFOOD), Public University of Navarre, Spain.}
\affil[5]{\small IdiSNA, Navarra Institute for Health Research, Navarra, Spain.}
\affil[6]{\small Biomedical Research Center Network on Physiopathology of Obesity and Nutrition, CIBEROBN, Institute of Health Carlos III, Madrid, Spain.}

\date{}

\makeatletter \pdfbookmark[0]{\@title}{title} \makeatother

\maketitle

\begin{abstract}
\noindent \textbf{Background:} Suicide remains a major public health concern worldwide, responsible for more than 700,000 deaths in 2021, accounting for approximately 1.1\% of all global deaths. While many high-income countries have reported declines in age-standardized suicide rates over the past two decades, recent evidence from Spain indicates increasing mortality among women, whereas suicide rates among men have remained relatively stable. To better understand these patterns and their potential underlying determinants, this study examines the spatial and temporal patterns of age-stratified suicide mortality across Spanish provinces from 2010 to 2022, with particular attention to sex-specific differences.
\\
\textbf{Methods:} Mixed Poisson models were applied to analyze provincial- and temporal-level suicide mortality rates, stratified by age and sex. The models accounted for spatial and temporal confounding effects and examined associations with various socioeconomic and contextual factors, including rurality and unemployment.\\
\textbf{Results:} Findings highlight the influence of rurality and unemployment on suicide mortality, with distinct gender-specific patterns. A 10$\%$ increase in the proportion of residents living in rural areas was associated with more than a 5$\%$ rise in male suicide mortality, while a 1$\%$ increase in the annual unemployment rate was linked to a 2.4$\%$ increase in female suicide mortality. Although male suicide rates remained consistently higher than female rates, a notable and steady upward trend was observed in female suicide mortality over the study period.\\
\textbf{Conclusions:} The use of sophisticated statistical models permits the detection of underlying patterns, revealing both geographic and temporal disparities in suicide mortality across Spanish provinces.
\end{abstract}


\bigskip

\section{Introduction} \label{sec:Intro}

According to the most recent report from the World Health Organization, more than 700,000 people died by suicide worldwide in 2021 \citep{world2025suicide},  accounting for approximately 1.1\% of all global deaths. The global age-standardized suicide rate was notably higher in males than in females (12.3 and 5.6 per 100,000 inhabitants, respectively) with the male-to-female suicide ratio reaching 3.2 in high-income countries. Although these countries show the highest age-standardized suicide rates, it is concerning that nearly 75\% of the total number of suicides occur in low- and middle-income countries.
These figures provide an overall picture, yet suicide rate trends differ markedly between regions. While suicide mortality has decreased in several parts of Europe, the Americas, and Australasia since the 2000s, countries such as the UK, Brazil, Mexico, the USA, South Korea, and Australia have experienced rising trends \citep{alicandro2019worldwide}.
%
Among these, the United States represents a notable example, with suicide remaining one of the leading causes of death and exhibiting a gradual increase in recent years.
Between 2020 and 2023, the national suicide rate increased by 4.5\%. In 2023 alone, more than 49,000 lives were lost to suicide, corresponding to an age-standardized rate of 14.1 per 100,000 inhabitants. Additionally, an estimated 12.8 million adults seriously considered ending their lives, 3.7 million made a plan, and 1.5 million attempted suicide \citep{CDC2025}.
In Europe, from 2000 to 2019, age-standardized suicide mortality rates declined by 2.3\% in females and 3.4\% in males \citep{ilic2022worldwide}. However, recent studies have reported some variation in suicide rates over the past decade in several countries, including Iceland, Norway, Sweden, and Spain \citep{zuin2025suicide}, although in many cases these changes are not statistically significant.

%
Beyond the pronounced sex differences, age is a well-recognized factor in suicide risk, with older adults generally exhibiting higher mortality rates across most countries \citep{WHO2014}. However, the relationship between age and suicide is complex and often non-linear, with specific age groups exhibiting higher vulnerability depending on the social context. This increased risk is especially notable among individuals suffering from physical illnesses, affective disorders and other mental disorders, conditions that frequently contribute to suicidal behavior \citep{conwell2011suicide, kim2025risk}.
Consistent with these patterns, estimates from the most recent Global Burden of Disease Study \citep{weaver2025global} indicate that the mean age at the time of death from suicide has increased over the past three decades, rising by four years among males (from 43.0 to 47.0) and by five years among females (from 41.9 to 46.9).

To better understand the dynamics and determinants of suicide mortality, numerous studies have examined its temporal evolution (often stratified by sex and age groups) or have sought to identify spatial patterns that highlight inequalities across subnational administrative units. Depending on their objectives and data availability, these studies typically adopt temporal, spatial, or spatio-temporal modelling frameworks.
When analysing temporal changes in national suicide mortality trends, researchers have employed a variety of methods, ranging from simpler statistical techniques such as moving average smoothing \citep{ivey2017suicide} and regression models for change-point analysis \citep{pompili2009suicide,hofstra2018springtime,ramalle2023trends}, to more advanced approaches like age-period-cohort models \citep{granizo1996age,riebler2012gender}, which are particularly useful for identifying long-term patterns.
Beyond purely temporal approaches, an increasing number of studies have investigated the spatial and spatio-temporal distribution of suicide mortality rates, aiming to identify regional disparities and examine the influence of socioeconomic factors on geographic disparities. Methodological approaches in this area range from descriptive mapping \citep{qi2014dynamic} to the application of inferential spatial models, including Poisson regressions with spatially structured random effects \citep{middleton2006geography,salmeron2013time} and the use of space-time scan statistics to detect potential high-risk clusters \citep{nunez2018trends}.

In this paper, we analyze annual suicide mortality in Spain from 2010 to 2022, using data disaggregated by sex and age group at the provincial level. This detailed stratification allows us to examine temporal trends and spatial distribution patterns across different population groups. Previous studies have reported a marked increase in suicide mortality among Spanish women between 2010 and 2016, while rates among men have remained relatively stable \citep{cayuela2020suicide}. The study period also includes the COVID-19 pandemic, which exacerbated several known risk factors for suicide and may have influenced changes in suicide mortality by gender, age group, and geographic region \citep{lantos2024impact}. Accordingly, this study provides an opportunity to evaluate the potential impact of this crisis on suicide mortality in Spain. Furthermore, several studies have highlighted pronounced geographical heterogeneity in suicide mortality across the country, with higher rates typically observed in regions with lower socioeconomic development \citep{santurtun2018does}.

To this end, we apply Poisson mixed models separately for each sex, incorporating structured random effects that borrow information from neighboring units to obtain reliable mortality rate estimates by effectively capturing spatial, temporal, and age-group dependencies \citep{goicoa2016age,goicoa2019flexible}.
Additionally, we include the percentage of the population living in rural areas, the unemployment rate, and the at-risk-of-poverty rate as explanatory variables to assess their influence on suicide mortality rates and to gain deeper insight into the factors driving geographic and temporal disparities.
These variables are modelled using an ecological regression framework to assess area-level associations while accounting for potential spatial confounding \citep{urdangarin2024simplified}. 
In summary, our research focuses on two main objectives: (i) analyzing sex-specific temporal and spatial patterns in age-stratified suicide mortality rates in Spanish provinces, 
and (ii) examining the influence of socioeconomic and contextual factors on suicide rates while accounting for potential confounding issues.

The rest of the paper is organized as follows. \autoref{sec:Data} provides a detailed description of the data. \autoref{sec:Methods} outlines the methodological framework employed for the data analysis and \autoref{sec:Results} presents the results. Finally, \autoref{sec:Discussion} ends with some conclusions. 

\section{Data}
\label{sec:Data}


This study uses suicide mortality counts and corresponding population data from the Spanish National Statistics Institute (INE) for the period 2010--2022. The data, which includes sex- and age-specific counts (ICD-10 codes X60--X84), is stratified by year and province across continental Spain. To assess potential associations with suicide mortality, we also include several socioeconomic and contextual variables. These consist of two spatial variables, the percentage of the population living in rural areas and the provincial unemployment rate, and two temporal variables, the national annual employment rate and the at-risk-of-poverty rate.
To construct a provincial-level rurality indicator, we use data from the Geographic Information System of the European Commission \citep{eurostat2024gisco}, which classifies local administrative units (municipalities in the case of Spain) into three categories of urbanisation: cities, towns or suburbs, and rural areas. Combining the 2020 classification with Spanish population data, we calculate the percentage of the population living in rural areas for each province. We obtain the remaining covariates (i.e., the provincial unemployment rate, national annual employment rate, and at-risk-of-poverty rate) from the National Statistics Institute.

In 2022, suicide remained the leading cause of external mortality in Spain, accounting for 22.8\% of all such deaths, ahead of \textit{unintentional threats to breathing} (22.1\%), \textit{accidental falls} (20.4\%), and \textit{transport accidents} (9.8\%) \citep{INE2023defunciones}.
Between 2010 and 2022, an average of 3,639 suicides were reported annually in Spain, representing 0.9\% of all deaths. The burden of suicide is highest among younger age groups. For individuals under 25, suicide accounts for 5.9\% of all deaths. This proportion rises to 11.7\% in the 25--44 age group, then decreases to 2.8\% for those aged 45--64 and 0.3\% for individuals aged 65 and over.
Our analysis focuses on the provinces of continental Spain, excluding the Balearic and Canary Islands and the autonomous cities of Ceuta and Melilla. A map of the provinces is provided in Supplementary \autoref{fig:Spain_provinces}. A total of 43,604 suicide deaths were recorded during this period, with 74.9\% correspond to males and 25.1\% to females. This distribution aligns with the pattern observed across the European Union, where males account for 76.7\% of all suicides \citep{Eurostat2024}.

\section{Methods} \label{sec:Methods}

Given the pronounced differences in suicide mortality rates between males and females in Spain, we fitted separate models for each sex. This approach allows for a more accurate representation of distinct age, spatial, and temporal trends. Because of these marked sex-specific dynamics, a shared spatial or temporal correlation structure could not be assumed across both sexes.

Initially, we aim to fit sex-specific models that incorporate structured random effects (including their interactions) for age group, province, and year, to estimate the spatio-temporal evolution of suicide mortality rates while sharing information across adjacent age groups.
We found that fully stratified age-space-time models exhibited excessive smoothing. This was likely due to the limited number of suicide cases within each stratum, a problem particularly pronounced among females. Such excessive smoothing, which hinders the models' ability to capture meaningful variation, is a documented limitation in the literature. For example, \cite{retegui2025prior} have shown that spatial analyses in very small areas can lead to this effect.
Fitting separate spatio-temporal models for each age group is also disadvantageous, as it fails to overcome data sparsity and prevents the model from leveraging shared patterns across age groups. To address these limitations and avoid over-smoothing, we decided to model the age-time and age-space dimensions separately, enabling a more reliable estimation of patterns within each component.

Given the structure of the age-space interaction models applied in \autoref{sec:AgeSpace_models}, we restricted the analysis to the provinces of continental Spain. Including the islands would have required artificially connecting them to the mainland in the spatial adjacency graph, which could have biased the spatial estimates by not reflecting true socioeconomic or health relationships.

\subsection{Age-time interaction models} \label{sec:AgeTime_models}

First, we analyze the data using a Bayesian hierarchical model that incorporates an age-time interaction effect. Let $O_{at}$ denote the observed number of suicide deaths in age group $a = 1, \dots, 8$ (corresponding to ages 10--19, ..., 70--79, and 80+), and time period $t=1,\dots,13$, representing years 2010 to 2022. Let $N_{at}$ and $\lambda_{at}$ denote the population at risk and the suicide mortality rate, respectively, for age group $a$ in year $t$. Conditional on the rates, we assume that the number of observed cases follows a Poisson distribution with mean $\mu_{at}=N_{at}\lambda_{at}$, that is
\begin{eqnarray*}
\setlength{\arraycolsep}{2pt}
\begin{array}{rcl}
O_{at}|\lambda_{at} & \sim & Poisson\left(\mu_{at}=N_{at} \lambda_{at} \right), \\[1.ex]
\log{\mu_{at}} & = & \log{N_{at}} + \log{\lambda_{at}}.
\end{array}
\end{eqnarray*}
Then, the log-rate is modelled as
\begin{equation}
\label{eq:Temporal_Model}
\log{\lambda_{at}} = \alpha + \phi_a + \gamma_t + \delta_{at},
\end{equation}
where $\alpha$ denotes the overall intercept (baseline log-rate), $\phi_a$ and $\gamma_t$ represents the main effects for age group and time, respectively, and $\delta_{at}$ denotes the age-time interaction term, capturing age-specific deviations from the additive structure of the marginal effects.
We evaluate alternative priors for the main effects of age and time to assess the model's sensitivity to prior specification.

The model performance led us to select a first-order random walk (RW1) prior for the age group random effect, ${\pphi} = (\phi_1, \ldots, \phi_8)^{'}$. We chose a RW1 prior for the temporal effect ${\ggamma} = (\gamma_1, \ldots, \gamma_{13})^{'}$ in the male model, but we applied a second-order random walk (RW2) prior for females. The RW2 prior introduces a higher degree of smoothness by borrowing strength from second-order temporal neighbors, which is particularly beneficial given the lower number of suicide cases observed among females. So, we assume that
\begin{equation*}
\pphi \sim N({\bf 0}, [\tau_{\phi} {\RR}_{\phi}]^{-}),
\qquad \mbox{and} \qquad
\ggamma \sim N({\bf 0}, [\tau_{\gamma} {\RR}_{\gamma}]^{-}),
\end{equation*}
where the symbol ${-}$ denotes the Moore-Penrose generalized inverse of a matrix, $\tau_{\phi}$ and $\tau_{\gamma}$ are precision parameters, and ${\RR}_{\phi}$ and ${\RR}_{\gamma}$ are structure matrices that define the precision structure of the RW priors, inducing smoothness across adjacent age groups and time periods, respectively.
Finally, we assume a multivariate normal prior for the age-time random effect, $\ddelta \sim N({\bf 0}, [\tau_{\delta} {\RR}_{\delta}]^{-})$ where ${\RR}_{\delta}$ is the precision matrix obtained as the Kronecker product (denoted by the symbol $\otimes$) corresponding to the four different types of interactions originally proposed by \cite{knorr2000}. A detailed summary of the prior assumptions for each random effect is provided in \autoref{tab:Priors}. See \cite{goicoa2018spatio} for a discussion on identifiability constraints in these models and their implementation using INLA.

\subsection{Age-space interaction models} \label{sec:AgeSpace_models}

Next, we analyze the data by fitting a Bayesian hierarchical model that incorporates an age-space interaction effect, allowing us to capture spatial variation in suicide mortality rates across provinces while accounting for age-specific differences. Let $O_{as}$ denote the observed number of suicide deaths in age group $a = 1, \dots, 8$ (corresponding to ages 10--19, ..., 70--79, and 80+), and small area $s=1,\dots,47$, representing the provinces of continental Spain. Let $N_{as}$ and $\lambda_{as}$ denote the population at risk and the suicide mortality rate, respectively, for age group $a$ in province $s$. As in the previous models, we assume that
\begin{eqnarray*}
\setlength{\arraycolsep}{2pt}
\begin{array}{rcl}
O_{as}|\lambda_{as} & \sim & Poisson\left(\mu_{as}=N_{as} \lambda_{as} \right), \\[1.ex]
\log{\mu_{as}} & = & \log{N_{as}} + \log{\lambda_{as}}.
\end{array}
\end{eqnarray*}
Then, the log-rate is modelled as
\begin{equation}
\label{eq:Spatial_Model}
\log{\lambda_{as}} = \alpha + \phi_a + \xi_s + \delta_{as},
\end{equation}
where $\alpha$ denotes the overall intercept (baseline log-rate), $\phi_a$ and $\xi_s$ represents the main effects for age group and space, respectively, and $\delta_{as}$ denotes the age-space interaction term.
For these particular set of models, we selected a RW1 prior distribution  for age-group random effect ${\pphi}=(\phi_1,\ldots,\phi_8)^{'}$ and an intrinsic conditional autoregressive (iCAR) prior for the spatial random effect ${\xxi}=(\xi_1,\ldots,\xi_{47})^{'}$. That is,
\begin{equation*}
\pphi \sim N({\bf 0}, [\tau_{\phi} {\RR}_{\phi}]^{-}),
\qquad \mbox{and} \qquad
\xxi \sim N({\bf 0}, [\tau_{\xi} {\RR}_{\xi}]^{-}),
\end{equation*}
where $\tau_{\phi}$ and $\tau_{\xi}$ are precision parameters, ${\RR}_{\phi}$ is the structure matrix of a RW1 prior, and the spatial structure matrix is defined as ${\RR}_{\xi}={\D}_W-{\W}$. Here, ${\W}=(w_{ij})$ denotes the spatial binary adjacency matrix (with $w_{ij} = 1$ if areas $i$ and $j$ share a common border, and $0$ otherwise), and ${\D}_W$ is a diagonal matrix whose entries correspond to the number of neighbours for each area. Again, four different prior specifications were considered for the age-space interaction effect. Further details are provided in \autoref{tab:Priors}.

\begin{table}[!t]
\renewcommand*{\arraystretch}{1}
\caption{\label{tab:Priors} Summary of the prior distributions on the age group, temporal, and spatial random effects, as well as on their respective interaction terms.}
\vspace{0.2cm}
\begin{tabular}{c|c|c}
\hline\\[-2ex]
& {\bf Age-time models} & {\bf Age-space models} \\[1.5ex]
& $\log{\lambda_{at}} = \alpha + \phi_a + \gamma_t + \delta_{at}$
& $\log{\lambda_{as}} = \alpha + \phi_a + \xi_s + \delta_{as}$ \\[1.5ex]
\hline & \multicolumn{2}{c}{} \\[-1.5ex]
Age effect (RW1) &
\multicolumn{2}{c}{$\pphi \sim N({\bf 0}, [\tau_{\phi} {\RR}_{\phi}]^{-})$} \\[1.5ex]
\hline & & \\[-1.5ex]
Temporal effect (RW1/RW2) &
$\ggamma \sim N({\bf 0}, [\tau_{\gamma} {\RR}_{\gamma}]^{-})$ & $-$ \\[1.5ex]
\hline & & \\[-1.5ex]
Spatial effect (iCAR) &
$-$ & $\xxi \sim N({\bf 0}, [\tau_{\xi} {\RR}_{\xi}]^{-})$ \\[1.5ex]
\hline & & \\[-1.5ex]
\multirow{7}{*}{Interaction effect}
& $\ddelta \sim N({\bf 0}, [\tau_{\delta} {\RR}_{\delta}]^{-})$ & $\ddelta \sim N({\bf 0}, [\tau_{\delta} {\RR}_{\delta}]^{-})$ \\[2ex]
& $\mbox{Type I:} \;\; {\RR}_{\delta}={\I}_{13} \otimes {\I}_{8},$
& $\mbox{Type I:} \;\; {\RR}_{\delta}={\I}_{47} \otimes {\I}_{8},$ \\[1.5ex]
& $\mbox{Type II:} \;\; {\RR}_{\delta}={\RR}_{\gamma} \otimes {\I}_{8},$
& $\mbox{Type II:} \;\; {\RR}_{\delta}={\RR}_{\xi} \otimes {\I}_{8},$ \\[1.5ex]
& $\mbox{Type III:}\;\; {\RR}_{\delta}={\I}_{13} \otimes {\RR}_{\phi},$
& $\mbox{Type III:}\;\; {\RR}_{\delta}={\I}_{47} \otimes {\RR}_{\phi},$ \\[1.5ex]
& $\mbox{Type IV:} \;\; {\RR}_{\delta}={\RR}_{\gamma} \otimes {\RR}_{\phi},$
& $\mbox{Type IV:} \;\; {\RR}_{\delta}={\RR}_{\xi} \otimes {\RR}_{\phi},$ \\[1.5ex]
\hline
\end{tabular}
\end{table}

\subsection{Ecological regression: addressing spatial and temporal confounding}
\label{sec:Spatial+}

Ecological regression models assess the relationship between area-level covariates and aggregated outcomes of interest, often within a Bayesian disease mapping framework. However, if spatial or temporal correlation between covariates and latent random effects is not adequately addressed, confounding may arise, leading to biased estimates and misleading conclusions.

The restricted spatial regression (RSR) method is probably one of the most widely adopted strategies to handle spatial confounding \citep{reich2006effects,hodges2010adding}, although several alternative approaches have been proposed. Recent studies, however, argue that RSR implicitly assumes the absence of unobserved covariates correlated with the observed ones, since it constrains the random effects to be orthogonal to the fixed effects \citep{gilbert2024}.
In contrast, the spatial+ approach proposed by \cite{dupont2022spatial+} mitigates spatial confounding by removing the spatial dependence of the covariates. In this study, we implement the simplified spatial+ method introduced by \cite{urdangarin2024simplified}, a modification that avoids separately fitting spatial models for the covariates to eliminate their spatial dependence. Below, we provide a brief overview of this method.

The simplified spatial+ approach begins by expressing each covariate ${\X}$ as a linear combination of the eigenvectors ${\U}_i$, $i=1,\ldots,n$, of the spatial precision matrix ${\RR}_{\xi}$. That is,
\begin{equation*}
{\X} = a_1 {\U}_1 + \ldots + a_n {\U}_n.
\end{equation*}
Next, the covariate is decomposed into two components, ${\X} = {\Z} + {\Z}^{*}$, where ${\Z}^{*}=a_{n-k}{\U}_{n-k} + \ldots + a_n{\U}_n$ contains the large-scale eigenvectors associated with the lowest eigenvalues, and ${\Z}=a_1{\U}_1 + \ldots + a_{n-(k+1)}{\U}_{n-(k+1)}$ comprises the remaining eigenvectors. Here, $k$ denotes the number of large-scale eigenvectors assigned to ${\Z}^{*}$, which is typically chosen not to exceed 20\% of the total eigenvectors.
Finally, assuming that collinearity between the fixed and random effects primarily arises from the eigenvectors associated with the lowest non-null eigenvalues, i.e., ${{\U}_{n-k}, \ldots, {\U}_{n}}$, the simplified spatial+ approach replaces the original covariate ${\X}$ with its spatially decorrelated component ${\Z}$ in the ecological regression model.
Note that the same approach can be applied to alleviate confounding when assessing associations with time-varying covariates in regression models that include temporally structured random effects.

\subsection{Model implementation}

The models described in the previous sections are implemented within a fully Bayesian framework using the integrated nested Laplace approximation (INLA) method \citep{rue2009approximate}, a widely used technique for approximate Bayesian inference in applied statistics \citep{rue2017bayesian}.
Specifically, we  fit the models using the novel hybrid approximation approach that combines the Laplace method with a low-rank Variational Bayes correction to the posterior mean \citep{van2023new,van2024low} using the \texttt{R-INLA} stable version 25.06.07 on R-4.5.1.

Regarding model hyperparameters, we assign improper uniform priors to all standard deviations (i.e., the square roots of the inverse precision parameters), while we use weakly informative normal priors with mean zero and precision 0.001 for the fixed effects.
To ensure meaningful comparison of precision parameters within each model, we scaled all structure matrices such that the geometric mean of their marginal variances equals one \citep{sorbye2014scaling}.

\section{Results} \label{sec:Results}

We first describe the main descriptive results of suicide mortality across sex, age groups, and provinces to provide context for the modelling analyses. We then report key results from the interaction models examining age, temporal, and spatial patterns in suicide mortality among males and females. Finally, we assess the influence of explanatory variables on the temporal trends and geographical distribution of these rates.

\subsection{Descriptive analysis}
\begin{figure}[!ht]
    \begin{center}
    \vspace{-0.2cm}
    \includegraphics[width=0.95\textwidth]{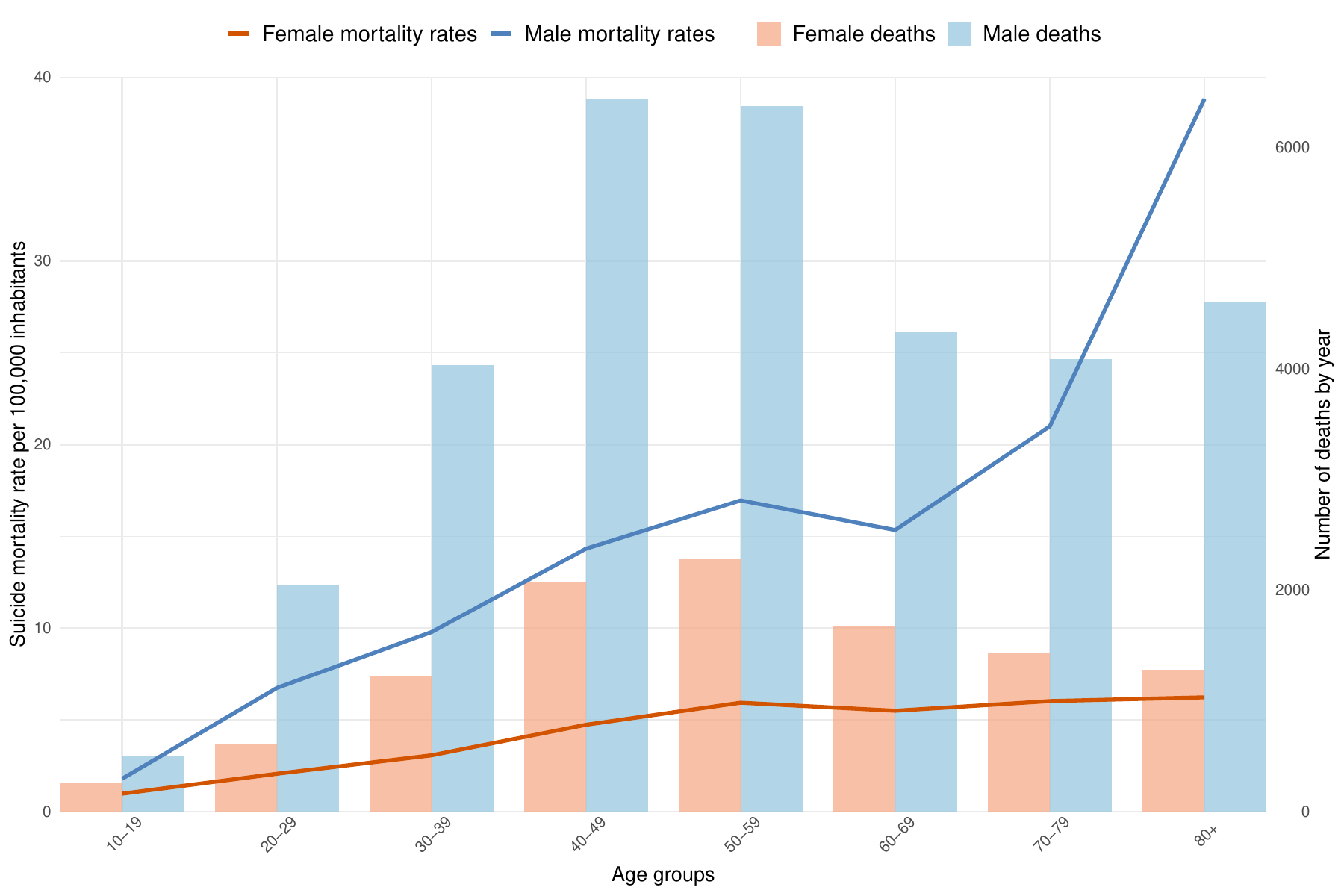}
    \end{center}
    \vspace{-0.2cm}
    \caption{Average annual number of suicide deaths and crude mortality rates by age group and sex in continental Spain, 2010--2022.}
    \label{fig:CrudeRates_SEXandAGE}
\end{figure}

In an initial exploratory analysis, we identified potential under-reporting in Madrid's data for the years 2010, 2011, and 2012. This finding aligns with explanations by some authors \citep{SANTURTUN2022150}, who note that until 2013, the INE (National Statistics Institute) lacked access to data from Madrid's Forensic Anatomical Institute.
Accordingly, these values were treated as missing and we then compute the national suicide mortality rates for those years by excluding both the population and mortality data from Madrid. Furthermore, when aggregating data into ten-year age intervals, suicide deaths in the youngest group (0--10 years) were negligible, and this group was therefore excluded from the analysis. \autoref{fig:CrudeRates_SEXandAGE} illustrates the average annual number of suicide deaths and crude mortality rates by ten-year age groups and sex in continental Spain from 2010 to 2022. The figure shows that males consistently exhibit higher mortality rates than females across all age groups. Over the study period, the average suicide mortality rate among males is 13.4 per 100,000 inhabitants, about three times higher than for females (4.3 per 100,000).
Age-specific trends reveal sex differences. In males, suicide mortality rates increase, peaking in the 50--59 age group before a slight decline in the 60--69 age group, followed by a steady rise with advancing age. Conversely, females show a similar peak around 50--59, but their rates remain relatively stable in subsequent age groups. These patterns underscore the strong influence of both sex and age on suicide mortality rates.

\begin{figure}[!ht]
    \begin{center}
    \includegraphics[width=\textwidth]{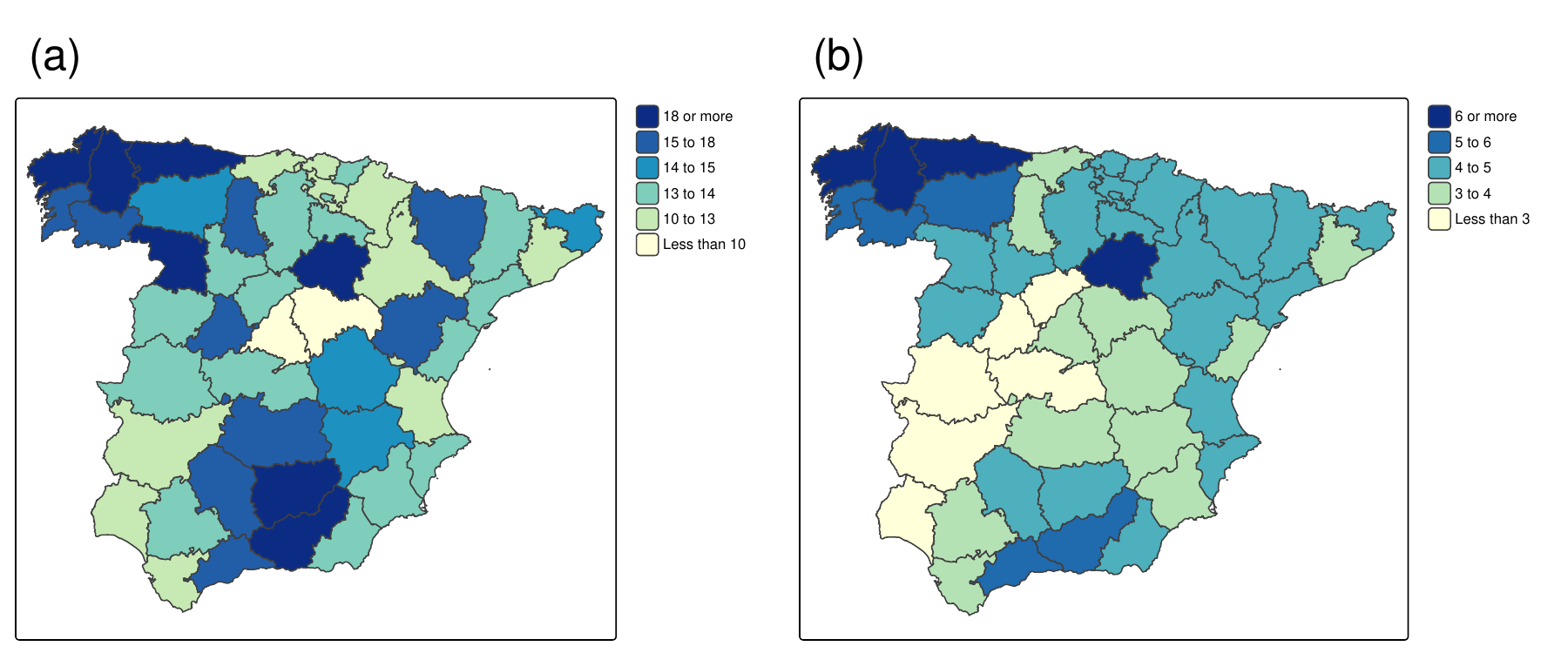}
    \end{center}
    \caption{Maps of suicide mortality rates (per 100,000 inhabitants) by province in continental Spain, 2010--2022. (a) Males; (b) Females.}
    \label{fig:CrudeRates_SEXandPROVINCE}
\end{figure}

Regarding the spatial distribution, notable differences in suicide mortality rates are evident across the provinces of continental Spain (see \autoref{fig:CrudeRates_SEXandPROVINCE}). While the northwestern regions shows the highest suicide mortality rates for both sexes, the geographic patterns in the two maps reveal marked differences, with female suicide rates displaying a smoother spatial pattern.
Among males, the province with the highest suicide mortality rate was Lugo (24.4 suicides per 100,000 inhabitants), with a rate nearly three times higher than that of the province with the lowest rate, Madrid (8.6 per 100,000). Among females, the disparity is even more pronounced, with Lugo reporting 8.5 suicides per 100,000 inhabitants, more than four times the rate observed in Segovia (2.1 per 100,000).
Assessing the geographical distribution of mortality rates by sex and age groups across regions is particularly challenging due to the inherent variability in crude rate estimates. However, the use of statistical models, such as those described in the following section, enables us to uncover underlying spatial patterns by smoothing rates across neighboring regions and age groups.

\begin{figure}[!ht]
    \begin{center}
    \vspace{0.5cm}
    \includegraphics[width=\textwidth]{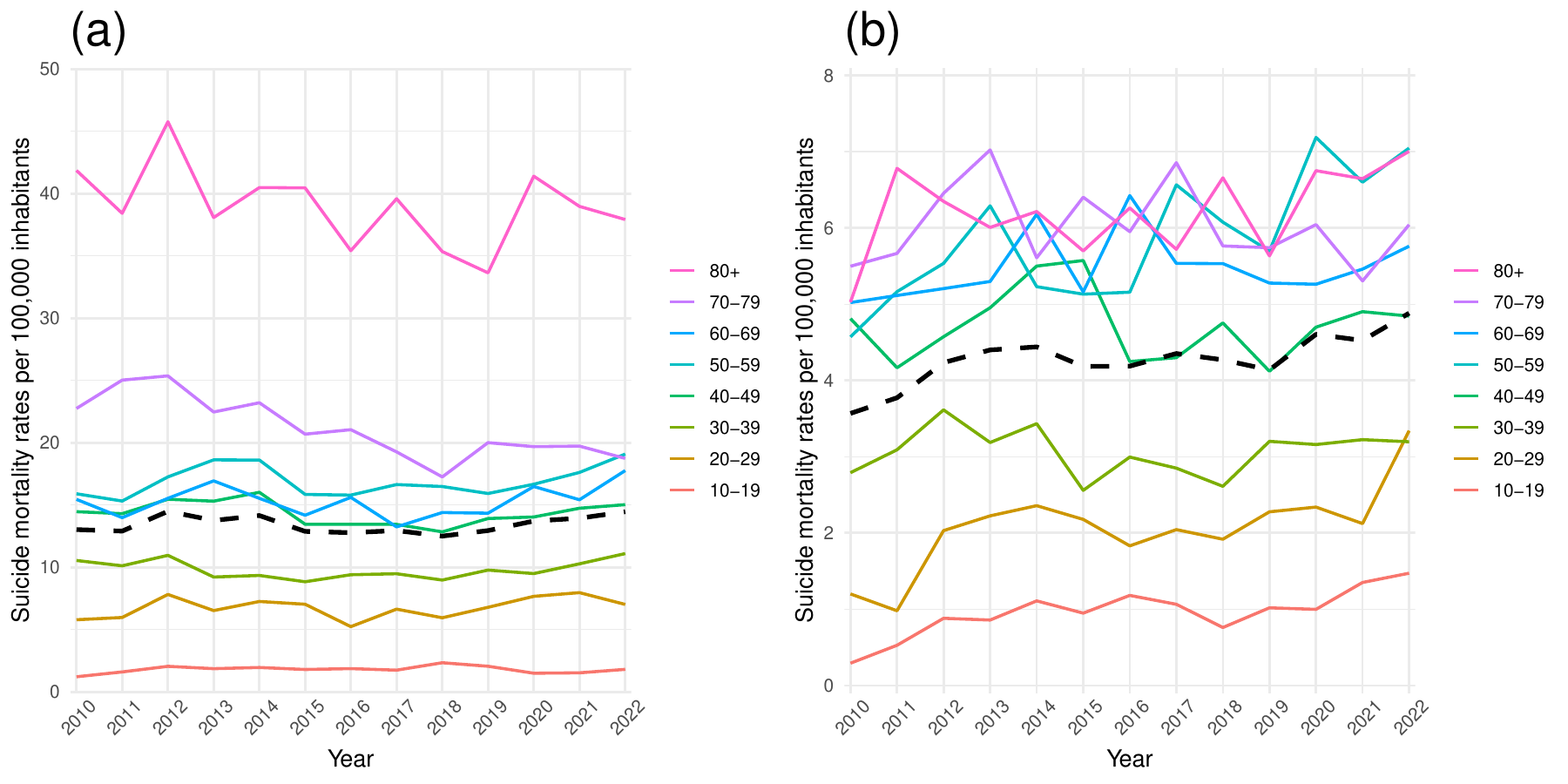}
    \end{center}
    \caption{Annual suicide mortality rates by sex and age groups. The dashed black line indicates the overall mortality rate. (a) Male population; (b) Female population.}
    \label{fig:CrudeRates_SEXandYEAR}
\end{figure}

The temporal evolution of annual suicide rates shows clear sex-based differences (see \autoref{fig:CrudeRates_SEXandYEAR}). Among males, most age groups do not exhibit a clear trend, a pattern also reflected in the overall male mortality rate. However, a slight downward tendency is observed in the older male age groups. In contrast, the overall mortality rate for females shows an average annual increase of approximately 2.4\%. Although age-specific rates among females display greater variability than those of males due to fewer observed cases, an upward trend is apparent across most female age groups. As with males, mortality rates remain very low for the youngest age groups.

\autoref{tab:CrudeRates_SEXandAgeandRurality} presents suicide mortality rates (with 95\% confidence intervals in brackets) disaggregated by sex, age groups, and rurality level. The rurality variable classifies the percentage of the population residing in rural areas within each province into three categories: low (less than 20\%), medium (20--40\%), and high (over 40\%). We compute confidence intervals using a normal approximation under the assumption that suicide counts follow a Poisson distribution. Across all age groups, suicide mortality rates are consistently higher in provinces with medium or high rurality levels, with the exception of female mortality in the 40--70 age group. These differences are particularly pronounced among males, with rates in highly rural areas exceeding those in low-rurality provinces by more than 30\%. In contrast, female rates remain relatively stable across rurality levels, except for a slight increase among those aged 70 and above.

\begin{table}[!ht]
\renewcommand{\arraystretch}{1.1}
\caption{Suicide mortality rates (per 100,000 inhabitants), with 95\% confidence intervals in brackets, by sex, age groups, and rurality level in Spanish provinces, 2010--2022.}
\label{tab:CrudeRates_SEXandAgeandRurality}
\begin{center}
\begin{tabular}{c|cccc}
\toprule
Age & Level of rurality & Males & Females \\
\hline
\multirow{6}{*}{$ \leq 40$ years}
& \multirow{2}{*}{Low ($<20\%$)}
  & 6.5        & 2.2        \\
& & [6.3, 6.7] & [2.0, 2.3] \\
& \multirow{2}{*}{Medium ($20-40\%$)}
  & 7.0        & 2.2        \\
& & [6.7, 7.4] & [2.4, 2.6] \\
& \multirow{2}{*}{High ($>40\%$)}
  & 6.6        & 2.6        \\
& & [5.8, 7.4] & [2.1, 3.1] \\
\hline
\multirow{6}{*}{$40-70$ years}
& \multirow{2}{*}{Low ($<20\%$)}
  & 14.7         & 5.2       \\
& & [14.4, 15.0] & [5.1, 5.4] \\
& \multirow{2}{*}{Medium ($20-40\%$)}
  & 17.2         & 5.8        \\
& & [16.7, 17.7] & [5.5, 6.1] \\
& \multirow{2}{*}{High ($>40\%$)}
  & 19.0         & 5.2        \\
& & [17.8, 20.1] & [4.6, 5.9] \\
\hline
\multirow{6}{*}{$ \ge 70$ years}
& \multirow{2}{*}{Low ($<20\%$)}
  & 25.8         & 5.9       \\
& & [25.1, 26.5] & [5.6, 6.2] \\
& \multirow{2}{*}{Medium ($20-40\%$)}
  & 31.3         & 6.6        \\
& & [30.0, 32.5] & [6.1, 7.1] \\
& \multirow{2}{*}{High ($>40\%$)}
  & 33.6         & 6.7        \\
& & [31.2, 36.0] & [5.8, 7.7] \\
\hline
\multirow{6}{*}{Total}
& \multirow{2}{*}{Low ($<20\%$)}
  & 12.6         & 4.2       \\
& & [12.5, 12.8] & [4.1, 4.2] \\
& \multirow{2}{*}{Medium ($20-40\%$)}
  & 15.1         & 4.6        \\
& & [14.8, 15.4] & [4.5, 4.8] \\
& \multirow{2}{*}{High ($>40\%$)}
  & 17.4         & 4.8        \\
& & [16.6, 18.1] & [4.4, 5.2] \\
\bottomrule
\end{tabular}
\end{center}
\end{table}

\subsection{Analysis of sex-specific temporal and spatial patterns}

\autoref{tab:DIC} presents the Deviance Information Criterion (DIC) and Watanabe-Akaike Information Criterion (WAIC) values used to assess and compare the fit of the age-time and age-space models separately for males and females. Both DIC and WAIC are measures of model fit that balance goodness-of-fit and model complexity: lower values indicate a better trade-off between how well the model explains the data and the effective number of parameters. In \autoref{tab:DIC}, the smallest values are highlighted in bold to indicate the preferred model for each scenario.
In line with the descriptive analysis, models including interaction effects consistently provide a better fit than their additive counterparts in all cases.
For age-time models, the Type II interaction model provides the best fit for male suicide rates according to the model selection criteria. This suggests that, beyond the main structured effects, the age-specific evolution during the study period shows a temporal pattern that does not necessarily have to be similar across consecutive age groups. In contrast, for female suicide rates we select the model incorporating a fully structured interaction random effect (i.e., Type IV model).
For the age-space models, we select the Type III interaction for both male and female data, indicating that the interaction effect captures non-spatial variability across provinces while preserving correlation between adjacent age groups.

\begin{table}[!t]
\begin{center}
\renewcommand*{\arraystretch}{1.2}
\caption{\label{tab:DIC} Comparison of age-time and age-space interaction models by sex, based on DIC and WAIC criteria. The smallest values are highlighted in bold to indicate the preferred model for each scenario.}
\vspace{0.2cm}
\begin{tabular}{c|c|cc|cc}
\hline
\multicolumn{2}{c|}{ } & \multicolumn{2}{c|}{\bf Age-time models} & \multicolumn{2}{c}{\bf Age-space models} \\[1.5ex]
\multicolumn{2}{c|}{ } & DIC & WAIC & DIC & WAIC \\[1.5ex]
\hline & & & & & \\[-2.5ex]
\multirow{5}{*}{\bf Males}
& Additive & 937.6 & 947.7 & 2828.1 & 2874.1 \\
& Type I   & 923.3 & 923.1 & 2693.3 & 2684.6 \\
& Type II  & {\bf 908.6} & {\bf 908.3} & 2703.0 & 2711.8 \\
& Type III & 925.8 & 932.4 & {\bf 2649.9} & {\bf 2639.6} \\
& Type IV  & 911.1 & 913.5 & 2658.0 & 2655.9 \\
\hline & & & & & \\[-2.5ex]
\multirow{5}{*}{\bf Females}
& Additive & 808.6 & 814.2 & 2145.6 & 2149.2 \\
& Type I   & 803.0 & 803.3 & 2146.5 & 2145.2 \\
& Type II  & 780.5 & 779.5 & 2144.5 & 2143.7 \\
& Type III & 806.3 & 812.2 & {\bf 2126.7} & {\bf 2115.0} \\
& Type IV  & {\bf 779.6} & {\bf 777.4} & 2126.4 & 2118.5 \\
\hline
\end{tabular}
\end{center}
\end{table}

Since the structure matrices for all random effects have been scaled, the posterior estimates of the precision parameter are comparable, allowing us to quantify the proportion of total variance attributable to each component.
In all models, the age group component accounts for the largest proportion of the variability, though the relative contributions of the remaining components vary. In the age-time interaction models, the main age effect explains 99.0\% of the variability for males and 96.0\% for females. After adjusting for the age effect, the temporal component accounts for 71.7\% of the remaining variability in males and 78.4\% in females, indicating similar heterogeneity in suicide trends across age groups for both sexes.
A similar pattern emerges in the age-space interaction models. Among males, age explains 93.9\% of the variability, space accounts for 4.7\%, and the interaction contributes only 1.4\%. In contrast, for females, the age component explains 83.4\%, while spatial and interaction effects contribute more substantially at 14.4\% and 2.3\%, respectively.

\begin{figure}[!ht]
    \begin{center}
    \includegraphics[width=0.95\textwidth]{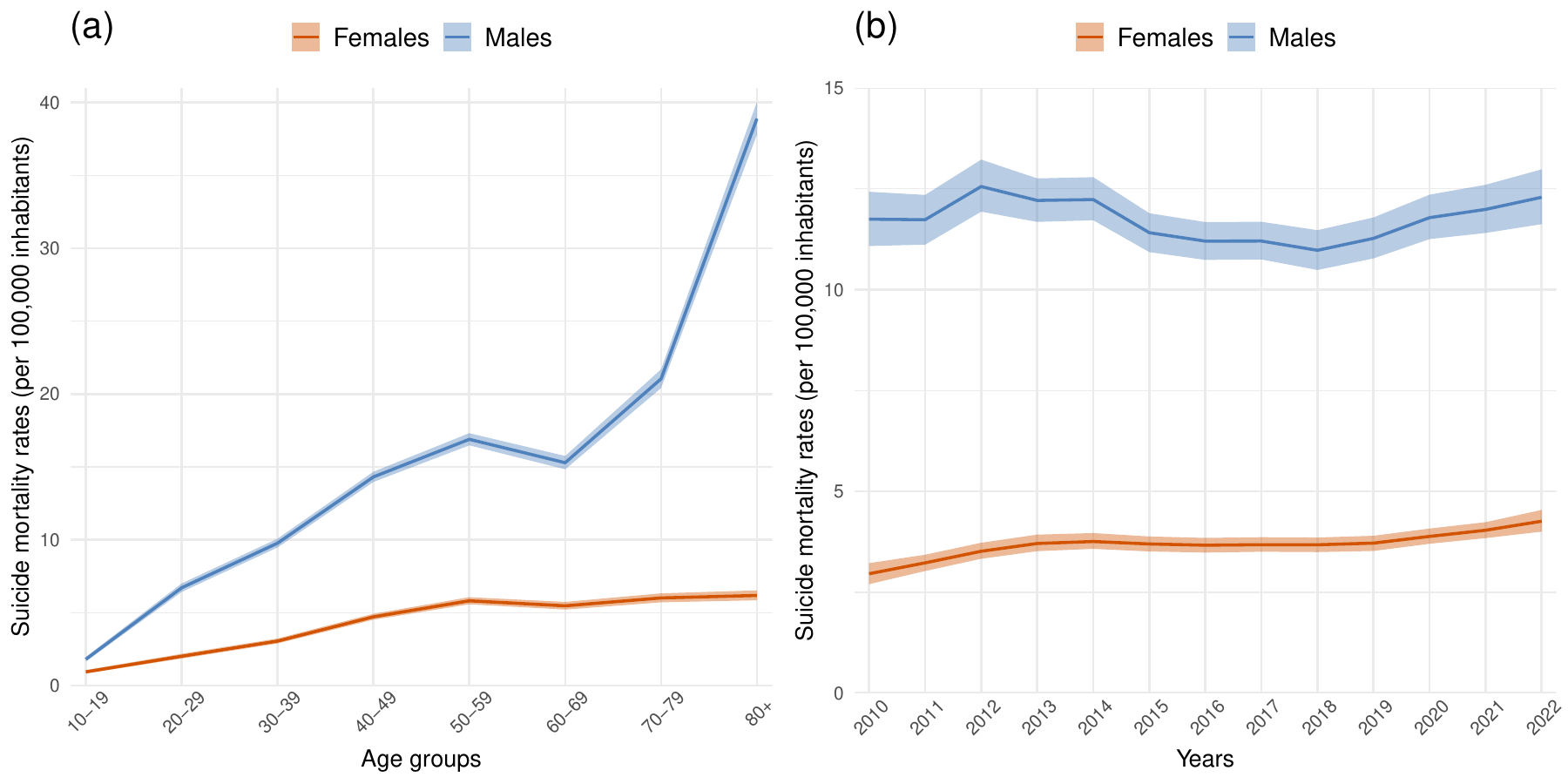}
    \end{center}
    \caption{Posterior median estimates and 95\% credible intervals of sex-specific age (a) and time (b) patterns of suicide mortality rates.}
    \label{fig:PosteriorPatterns_SEXandYEAR}
\end{figure}

\autoref{fig:PosteriorPatterns_SEXandYEAR} shows the posterior median estimates and 95\% credible intervals for the sex-specific age and time patterns of suicide mortality rates, calculated as $\exp(\alpha + \phi_a)$ and $\exp(\alpha + \gamma_t)$, respectively. These model-based trends closely resemble the aggregated crude rates from the descriptive analysis, effectively capturing the main age-specific and temporal variations along with their associated uncertainty.
Notably, the smoothed estimates indicate overall suicide mortality rates of approximately 12.0 and 3.7 deaths per 100,000 inhabitants for males and females, respectively, slightly lower than the corresponding crude rates.

\begin{figure}[!ht]
    \begin{center}
    \includegraphics[width=\textwidth]{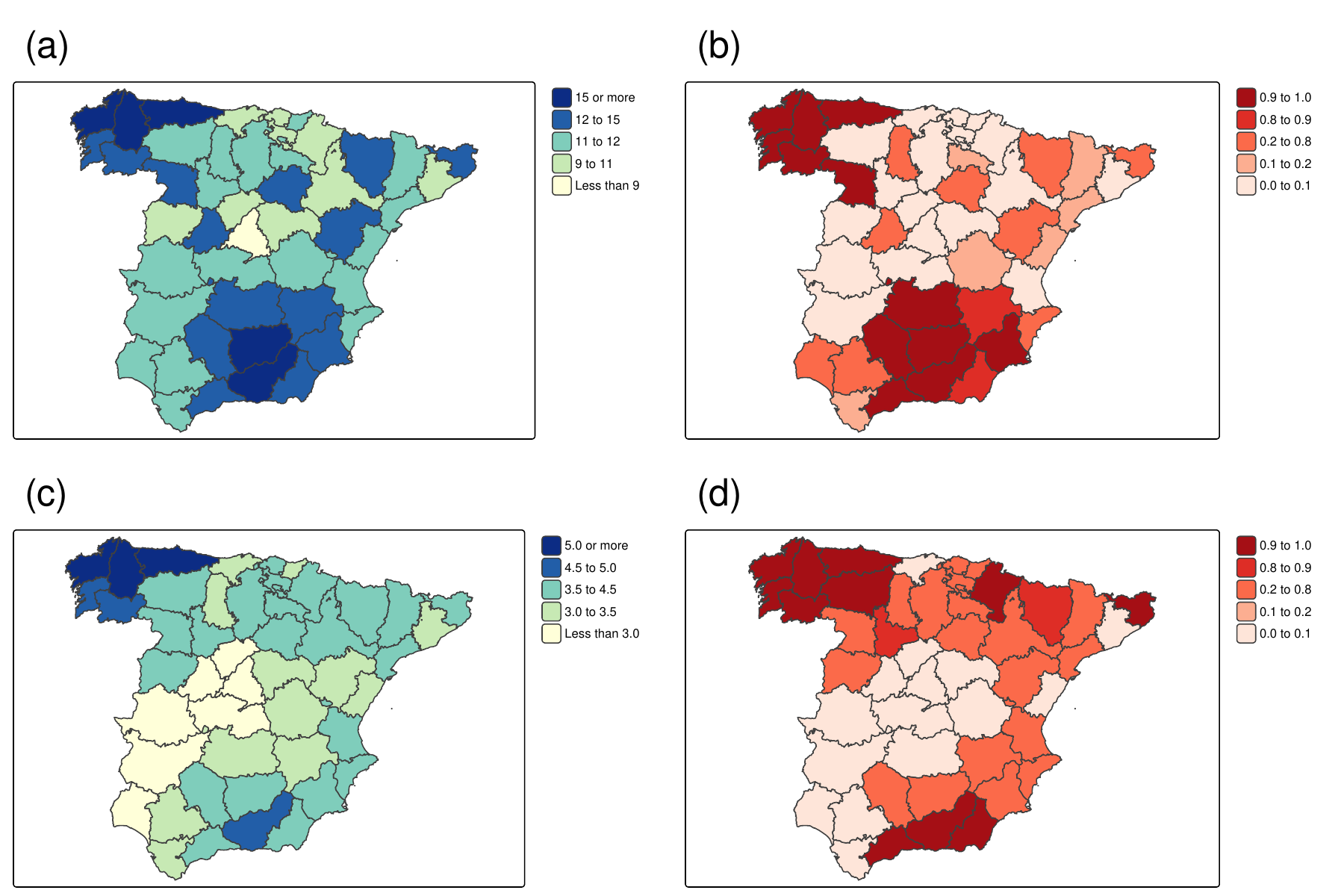}
    \end{center}
    \caption{Sex-specific spatial patterns of suicide mortality rates. (a) Posterior median estimates for males; (b) Posterior exceedence probabilities for males; (c) Posterior median estimates for females; (d) Posterior exceedence probabilities for females.}
    \label{fig:PosteriorPatterns_SEXandPROVINCE}
\end{figure}

Similarly, the left column in \autoref{fig:PosteriorPatterns_SEXandPROVINCE} displays maps of the posterior median estimates for the sex-specific spatial patterns of suicide mortality rates, computed as $\exp(\alpha + \xi_s)$. The right column shows the posterior exceedence probabilities $Pr(\xi_s>0 | {\bf O})$, highlighting regions with a high probability of rates above the overall national mortality rate.
Once again, the maps of estimated rates can be interpreted as smoothed representations of the aggregated crude rates presented in the descriptive analysis. For males, greater spatial heterogeneity is evident across provinces, whereas the female map reveals a more gradual and consistent spatial gradient. Notably, the posterior exceedance probability maps identify regions in both the north and the south with rates significantly above the overall mean.

\begin{figure}[!ht]
    \begin{center}
    \includegraphics[width=\textwidth]{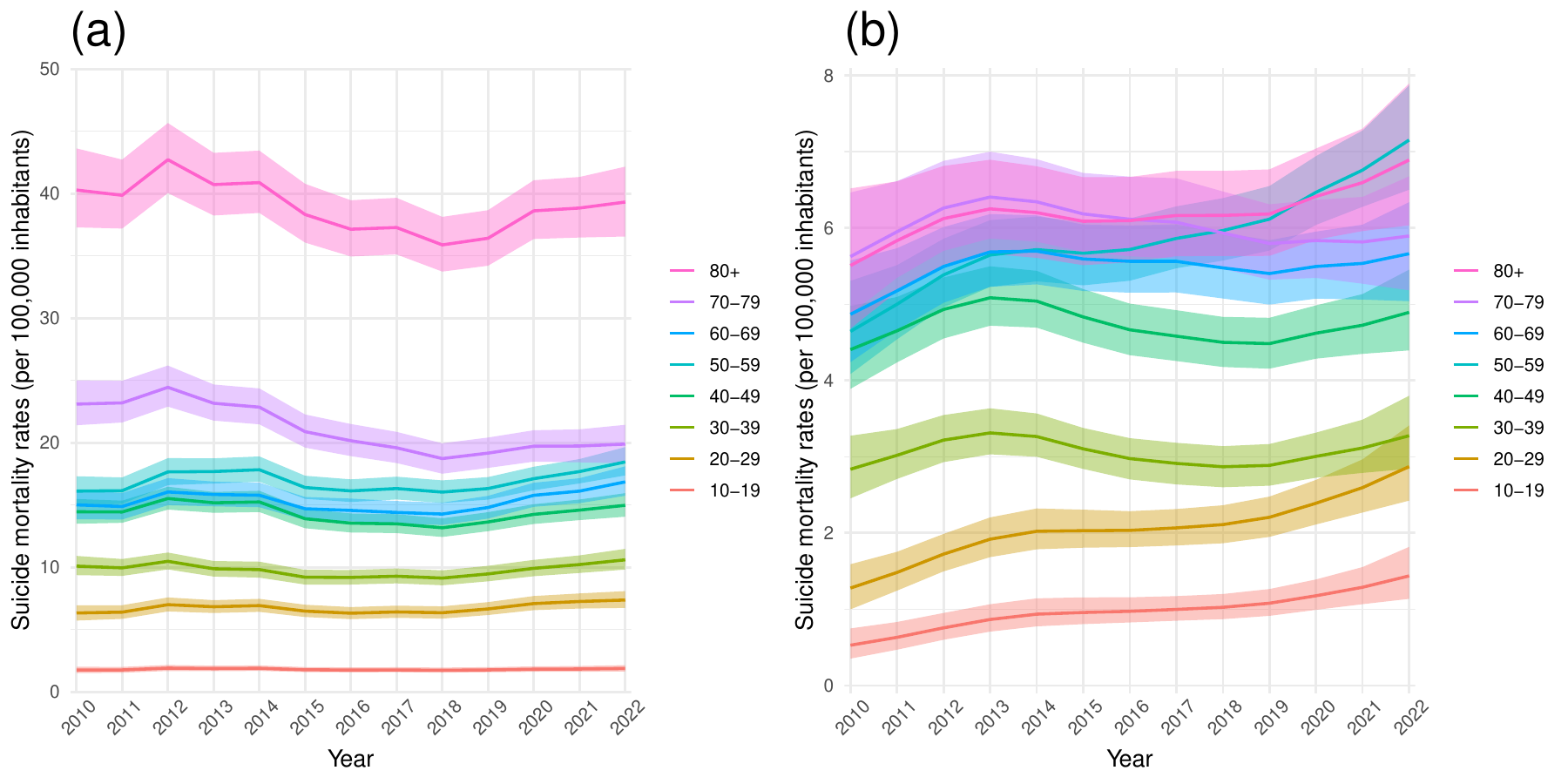}
    \end{center}
    \vspace{-0.2cm}
    \caption{Temporal evolution of posterior median estimates and 95\% credible intervals of suicide mortality rates by age group. Different scales are used for the Y axis to improve the visualization of the rates. (a) Male population; (b) Female population.}
    \label{fig:EstimatedRisks_AgeTime}
\end{figure}

The temporal evolution of sex-specific suicide mortality rate estimates by age group is presented in \autoref{fig:EstimatedRisks_AgeTime}.  As expected, the credible intervals for female estimates are substantially wider, reflecting that the number of recorded deaths is approximately three times higher in males than in females across most age groups.
In general, suicide mortality rates progressively increase with age, except among males aged 40--70 and females over 50, where no significant differences are observed. Over the study period, temporal trends shows a slight decline among elderly males, whereas for females in these age groups (and in most others) the trend clearly increases.

\begin{figure}[!ht]
    \begin{center}
    \includegraphics[width=\textwidth]{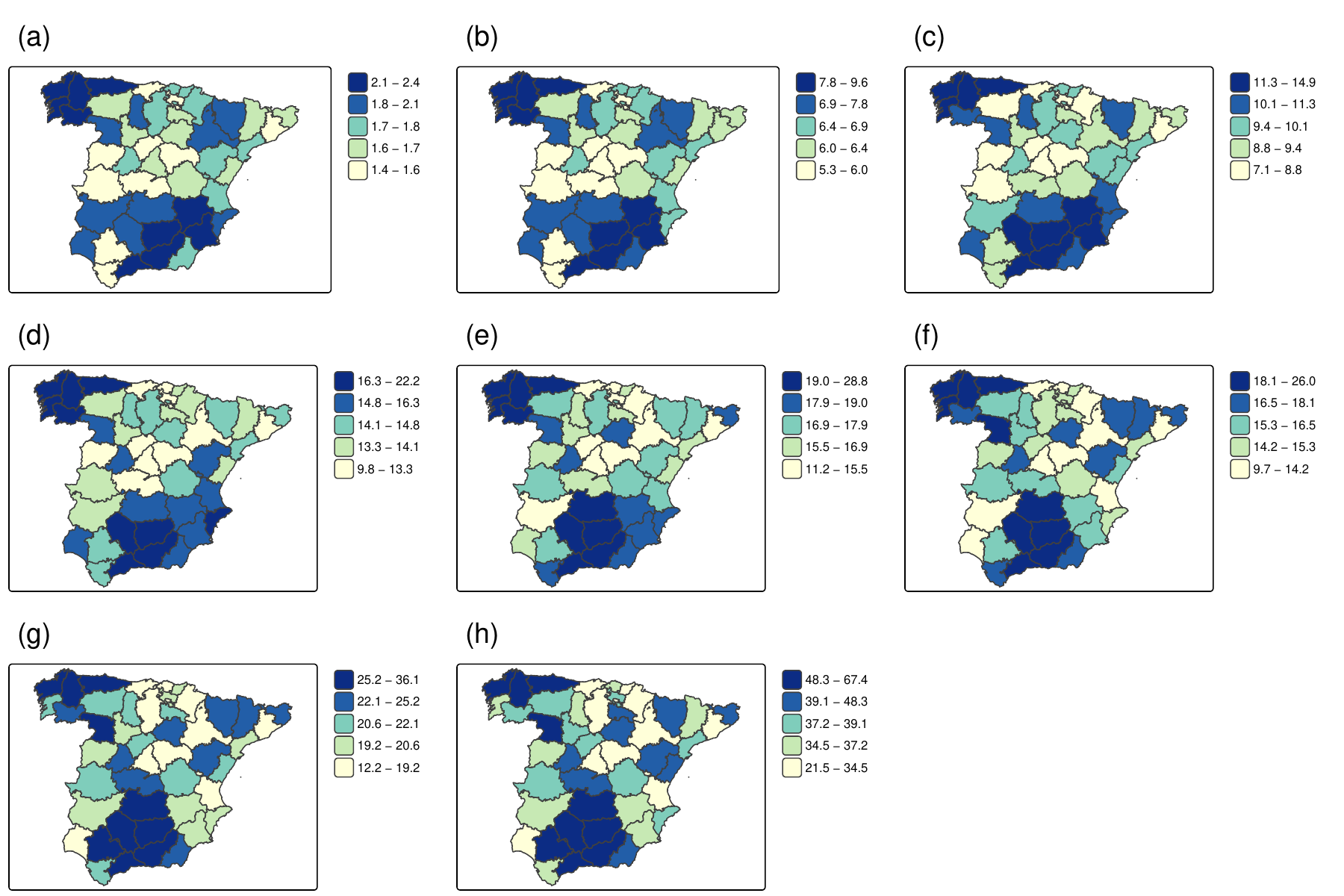}
    \end{center}
    \caption{Maps of posterior median estimates of male suicide mortality rates by age group. The breaks in the color scales were defined using the quintiles of the estimated rates within each age group. (a) Age group 10--19; (b) Age group 20--29; (c) Age group 30--39; (d) Age group 40--49; (e) Age group 50--59; (f) Age group 60--69; (g) Age group 70--79; (h) Age group 80+.}
    \label{fig:EstimatedRisks_AgeSpace_Males}
\end{figure}

\begin{figure}[!ht]
    \begin{center}
    \includegraphics[width=\textwidth]{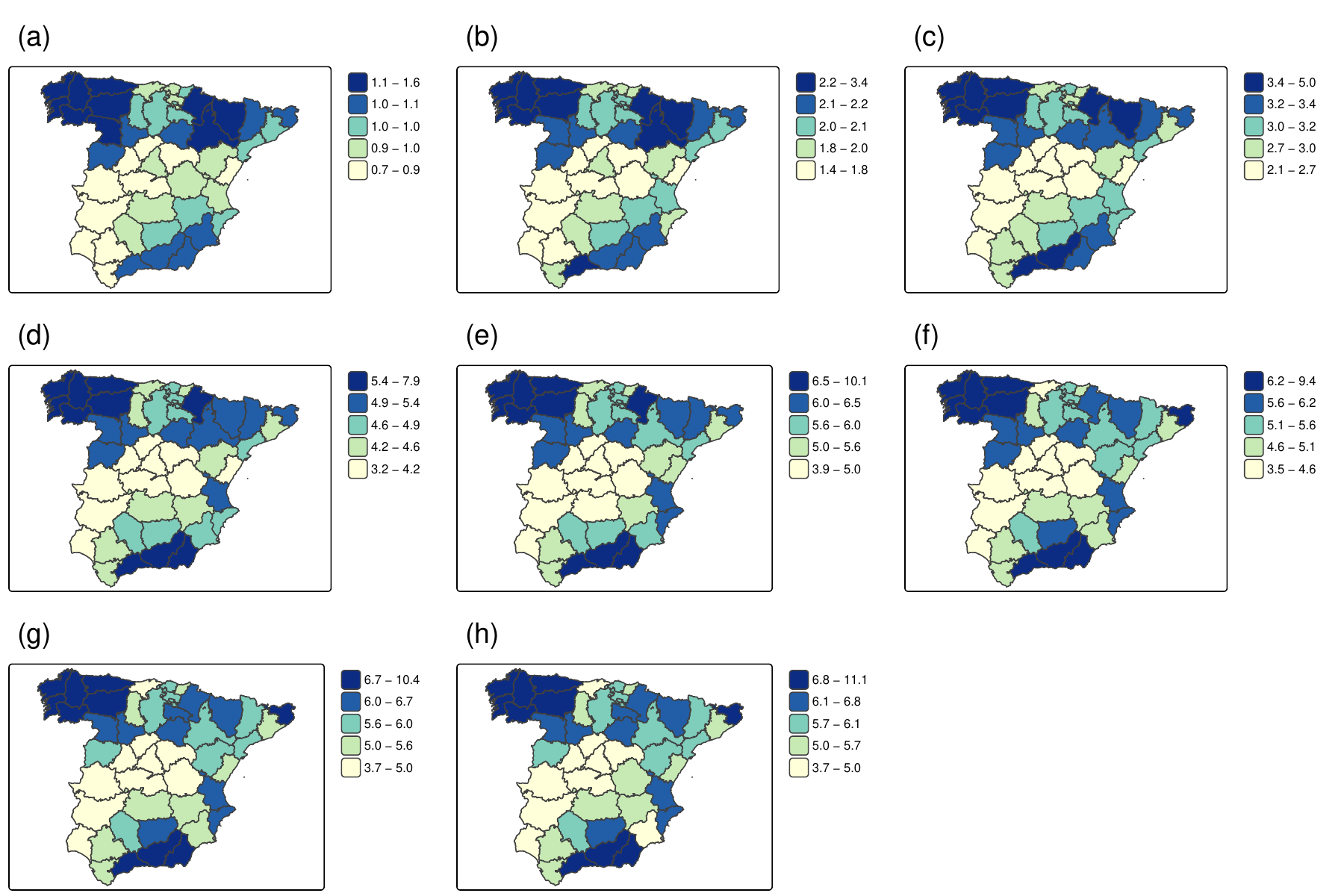}
    \end{center}
    \caption{Maps of posterior median estimates of female suicide mortality rates by age group. The breaks in the color scales were defined using the quintiles of the estimated rates within each age group. (a) Age group 10--19; (b) Age group 20--29; (c) Age group 30--39; (d) Age group 40--49; (e) Age group 50--59; (f) Age group 60--69; (g) Age group 70--79; (h) Age group 80+.}
    \label{fig:EstimatedRisks_AgeSpace_Females}
\end{figure}

Finally, the maps illustrating the geographical distribution of suicide mortality rates by age group are shown in \autoref{fig:EstimatedRisks_AgeSpace_Males} (males) and \autoref{fig:EstimatedRisks_AgeSpace_Females} (females). For visualization purposes, the breaks in the color scale were defined using the quintiles of the estimated rates within each age group, so comparisons across panels should be interpreted with caution.
In both cases, the maps reveal a consistent geographical pattern across age groups, with overlapping areas of high and low rates. Nonetheless, some differences emerge between specific age groups.
Among the male population, provinces in western and southern inland Spain (including Galicia, much of Castilla-La Mancha, and the interior of Andalusia) consistently show higher mortality rates in the older age groups, particularly from 50--59 years onward, with rates especially pronounced in the 70--79 and 80+ age groups. In contrast, some provinces in the north, northeast, and along the Mediterranean coast (such as the Basque Country, Navarra, Arag{\'o}n, and the Valencian Community) tend to exhibit lower mortality rates within the same older age groups.
Among the female population, the underlying geographic pattern exhibits stronger spatial dependence than in males. In particular, most regions in the northwest of Spain show consistently high mortality rates across all age groups. Meanwhile, some northern provinces (such as Navarra and Arag{\'o}n) also display relatively high mortality rates in younger age groups, but a downward trend emerges as age increases. In contrast, certain southern regions exhibit the opposite pattern: mortality rates there not only remain above the national average but also show a clear increasing trend among older age groups.

We also calculate the posterior exceedance probabilities $Pr(\lambda_{as}>\hat{\alpha}+\hat{\phi}_a | {\bf O})$, which quantify the probability that area-level rates exceed the overall mean within each age group, thereby accounting for the uncertainty associated with the posterior marginal estimates. Maps of these probabilities are presented in \autoref{fig:ExceedanceProbs_AgeSpace_Males} and \autoref{fig:ExceedanceProbs_AgeSpace_Females} of the supplementary material.

\subsection{Analysis of socioeconomic and geographic disparities}

\begin{figure}[!ht]
    \begin{center}
    \vspace{0.5cm}
    \includegraphics[width=\textwidth]{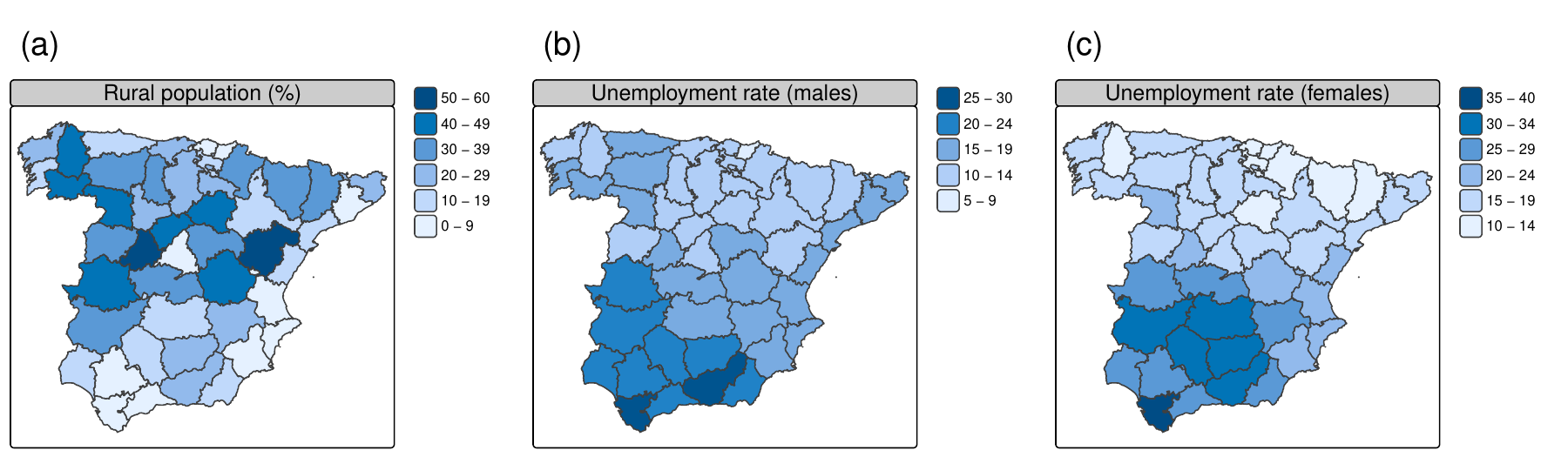}
    \end{center}
    \caption{Spatial distribution of percentage of the population living in rural areas (a) and unemployment rate for males (b) and females (c).}
    \label{fig:Covariates_Spatial}
\end{figure}

\begin{figure}[!ht]
    \begin{center}
    \includegraphics[width=\textwidth]{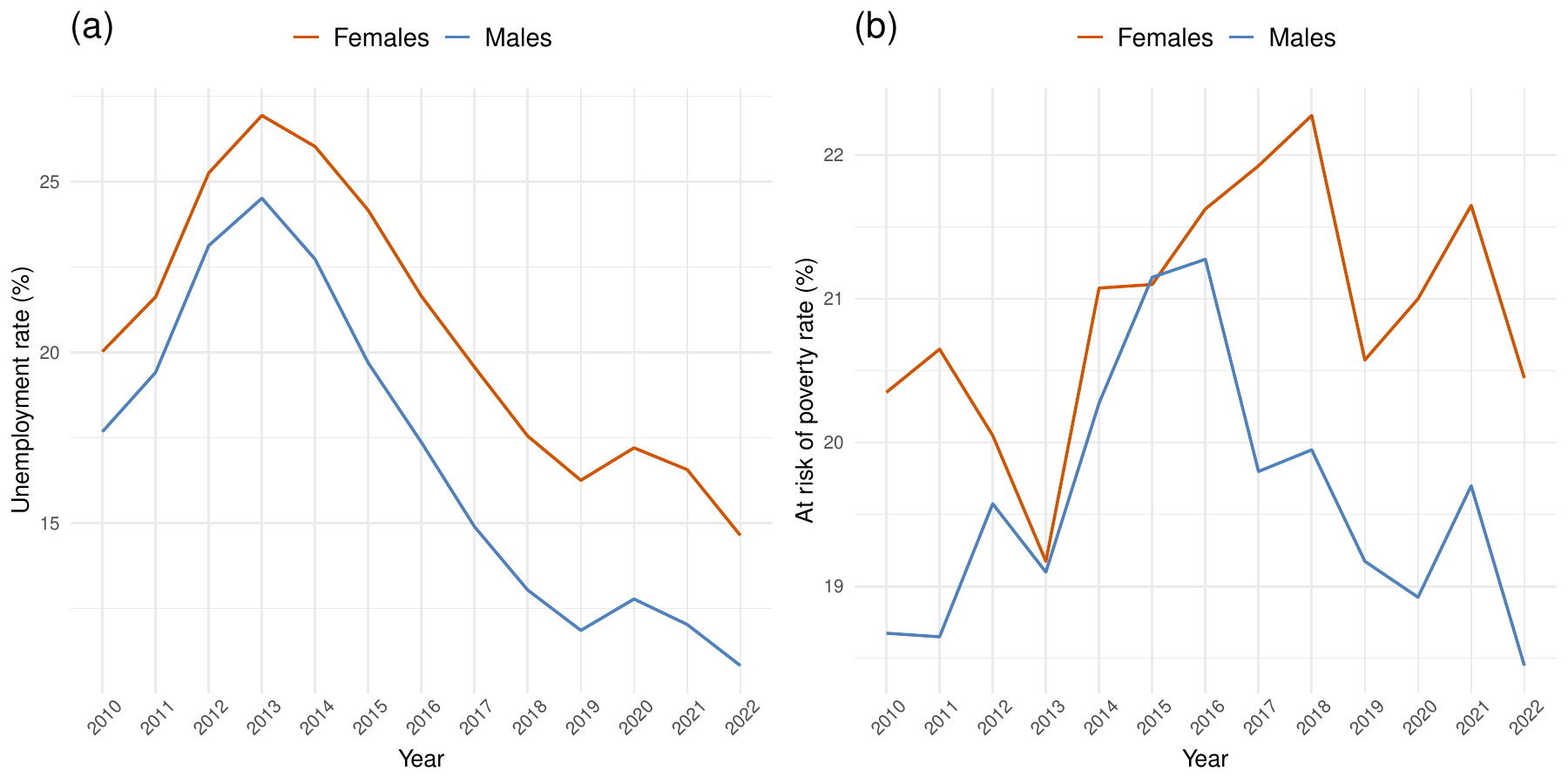}
    \end{center}
    \caption{Sex specific temporal trends in the annual unemployment rate (a) and the at-risk-of-poverty rate (b).}
    \label{fig:Covariates_Temporal}
\end{figure}

After examining the sex-specific temporal and spatial patterns of suicide mortality across age groups, this section examines whether socioeconomic and contextual factors help account for the observed geographical and temporal disparities in suicide rates.
\autoref{fig:Covariates_Spatial} illustrates the spatial distribution of the percentage of the population living in rural areas and the average unemployment rate between 2010 and 2022, presented separately for males and females. Similarly, \autoref{fig:Covariates_Temporal} shows the sex-specific temporal trends in the annual employment rate and at-risk-of-poverty rate.

As described in \autoref{sec:Spatial+} the simplified spatial+ approach was applied to obtain ``unconfounded'' fixed-effect estimates for the covariates of interest. The same prior structures for the random effects as in the final selected models were used for both the age-space and age-time interaction models.
To obtain the decorrelated components for the spatial covariates, we removed the $k=5$ large-scale eigenvectors associated with the spatially structured precision matrix, whereas for the temporally varying covariates, only $k=2$ eigenvectors of the temporally structured precision matrix were removed. In both cases, the covariates were standardized, ensuring that the estimated regression coefficients are directly comparable in terms of relative effect sizes and reducing potential issues arising from differences in variable scales.

\autoref{tab:Covariates} presents the posterior estimates of the fixed effects obtained under the simplified spatial+ approach. Among males, the proportion of the population living in rural areas shows a statistically significant positive association with suicide mortality, as the corresponding 95\% credible interval excludes zero. In contrast, no clear effects are observed for the provincial unemployment rate in males or for either covariate in females when using the age-space interaction models. In the age-time interaction models, the annual unemployment rate exhibits a significant positive effect for females, whereas for males the evidence is weaker, with the 95\% credible interval narrowly overlapping zero. Finally, the at-risk-of-poverty rate shows no significant association with suicide mortality in either sex.

\begin{table}[!ht]
\begin{center}
\caption{\label{tab:Covariates} Parameter estimates (posterior median, 2.5\% and 97.5\% quantile) for models fitted under the simplified spatial+ approach.}
\vspace{0.2cm}
\begin{tabular}{l|rrr|rrr}
\hline\\[-2ex]
\multicolumn{1}{c|}{\bf Age-space} & \multicolumn{3}{c|}{Results for males} & \multicolumn{3}{c}{Results for females} \\[1.5ex]
{\bf interaction models} & $2.5\%$ & Median & $97.5\%$ & $2.5\%$ & Median & $97.5\%$ \\[1ex]
\hline\\[-2ex]
Rural population  & 0.033 & 0.079 & 0.125 & -0.070 & -0.004 & 0.060 \\[1ex]
Unemployment rate &-0.088 & 0.011 & 0.108 & -0.226 & -0.083 & 0.057 \\[1ex]
\hline
\hline & & & & & & \\[-1.5ex]
\multicolumn{1}{c|}{\bf Age-time} & \multicolumn{3}{c|}{Results for males} & \multicolumn{3}{c}{Results for females} \\[1.5ex]
{\bf interaction models} & $2.5\%$ & Median & $97.5\%$ & $2.5\%$ & Median & $97.5\%$ \\[1ex]
\hline\\[-2ex]
Unemployment rate       & -0.001 & 0.069 & 0.140 & 0.018 & 0.093 & 0.172 \\[1ex]
At-risk-of-poverty rate & -0.043 &-0.014 & 0.021 &-0.031 & 0.000 & 0.030 \\[1ex]
\hline
\end{tabular}
\end{center}
\end{table}

To aid interpretation, the fixed-effect estimates were transformed back to the original scale and expressed as rate ratios. For males, a 10 percentage point increase in the proportion of the population living in rural areas is associated with an estimated 5.4\% increase in suicide mortality (rate ratio: 1.054; 95\% CrI: $[1.022, 1.086]$). Among females, we did not detect a significant association, as the rate ratio remains close to one (0.997; 95\% CrI: $[0.955, 1.040]$) with wide uncertainty.
Regarding the age-time interaction models, a one-percentage-point increase in the annual unemployment rate is associated with a 2.4\% increase in female suicide mortality (rate ratio: 1.024; 95\% CrI: $[1.004, 1.045]$), whereas the corresponding effect for males is weaker and more uncertain (rate ratio: 1.015; 95\% CrI: $[1.000, 1.031]$).

\section{Discussion} \label{sec:Discussion}

This study provides a comprehensive analysis of suicide mortality in Spain, highlighting notable differences across sex, age group, and geographic region.
Our definition of suicide mortality is limited to ICD-10 codes X60--X84. While other codes could potentially be considered suicides (Y87: 'Sequelae or late effects of intentional self-harm, assault, and events of undetermined intent', Y10--Y19: 'Poisoning of undetermined intent', and Y20--Y34: 'Other events of undetermined intent'), the exclusion of these codes may result in a very small undercount of suicide deaths. However, their contribution is minimal (accounting for less than 2\% of total cases during the study period) and is unlikely to materially affect the spatial or temporal patterns observed in our study.

Although male suicide rates consistently exceeded those of females between 2010 and 2022, we observed a pronounced upward trend  among females. This increase was particularly evident in younger age groups, with posterior median estimates indicating a 173\% rise among those aged 10--19 and a 124\% rise among those aged 20--29, as well as a 54\% increase among females aged 50--59. In terms of spatial patterns, males exhibited greater heterogeneity across provinces, with a larger number of areas showing significantly elevated rates. In contrast, females displayed a smoother spatial distribution. Nonetheless, two regions -Galicia in the northwest and southern Andalusia- were consistently identified in both sexes as having mortality rates markedly above the national mean. Additionally, Navarra and its neighboring eastern provinces displayed particularly elevated suicide rates among younger females.


Our findings also show a marked increase in suicide mortality among the oldest male age groups in Spain, particularly from age 80 onwards, a pattern that aligns with evidence reported in the international literature. Previous research has demonstrated that suicide mortality rises progressively with age and reaches its highest levels among older men, including those aged 85-90 years \citep{shah2016suicide}. Other studies similarly report especially elevated rates among individuals aged 80 years and older \citep{naghavi2019global}, and evidence suggests that this trend may even persist among centenarians \citep{de2022late}.
Physical illness and functional disability, both highly prevalent in late life, may help explain the particularly elevated suicide rates observed among the oldest men. In Spain, mental health organizations have highlighted unwanted loneliness as an additional risk factor in this population. Physical health problems and functional limitations often entail loss of autonomy, chronic pain, social withdrawal, and depressive symptoms, all of which have been identified as important contributors to suicide risk in older adults \citep{shah2016suicide}. Furthermore, some studies suggest that the impact of physical health problems on suicidal behaviour may vary across advanced age groups. For example, a systematic review reported that hospitalization and certain chronic diseases, such as cancer, were significant predictors of suicide among men aged 80 years and older, but not among those aged 70-79 \citep{fassberg2016systematic}.

When examining the influence of socioeconomic and contextual factors, we found that suicide rates were positively associated with the proportion of the population living in rural areas among males, and with the annual unemployment rate among females. In contrast, provincial unemployment and at-risk-of-poverty rates showed no clear association with suicide mortality in either sex.
The observed association between male suicide rates and rurality is consistent with previous studies conducted in Ireland \citep{o2016area}, Germany \citep{helbich2017spatiotemporal}, the USA \citep{fontanella2015widening,ivey2017suicide} and England \citep{middleton2006geography,gascoigne2025spatio}, among others. This pattern is further supported by a recent systematic review and meta-analysis \citep{davico2025relationship}, which demonstrated that lower population density, commonly used as a proxy for rurality, was associated with higher suicide rates in the majority of countries examined.
However, the positive association between annual unemployment and female suicide rates is somewhat paradoxical, as previous studies in Spain and other countries have generally reported stronger associations among males \citep{cunningham2022gendered, blazquez2023associations}. Interestingly, \cite{kposowa2019new} observed a similar pattern in the United States, where unemployed women exhibited higher suicide rates than unemployed men, consistent with our findings and suggesting that gender-specific vulnerabilities may underlie this relationship.

Considering the multifactorial nature of suicide risk, arising from the interplay of biological, psychological, clinical, and socio-environmental factors \citep{turecki2019suicide}, Bayesian hierarchical models with structured random effects offer a powerful framework to capture complex spatial, temporal, and age-related patterns in mortality rates. These models enable comprehensive uncertainty quantification and incorporate correlations across provinces, time periods, and age groups, resulting in more reliable and robust estimates.
Another key strength of our work is the use of the simplified spatial+ approach within an ecological regression framework, which accounts for potential spatial and temporal confounding between fixed and random effects, thereby reducing the risk of biased fixed-effect estimates that could result if these dependencies were ignored.

One limitation of our study is that attempts to fit fully stratified age-space-time models led to excessive smoothing. This is because the data was too sparse at this level of resolution, which caused excessive smoothing and could have led to incorrect interpretations \citep{retegui2025prior}.
The analysis of suicide mortality is further limited by the potential underreporting and regional inconsistencies in registry data, largely due to the challenging nature of cause-of-death classification. For instance, official suicide rates in Spain have historically been lower than those in some other European countries \citep{snowdon2021spain}. Other researchers \citep{SANTURTUN2022150} posit that this discrepancy may be attributed, in part, to the fact that Spain's Statistical Registry of Deaths with Judicial Intervention was previously filled out by a civil servant, who often lacked adequate information about the cause of death, rather than a forensic doctor.

Furthermore, the absence of monthly suicide mortality data at the subnational level prevented an examination of seasonal patterns, such as the peak in suicide rates during spring and summer that has been previously documented in Spain \citep{salmeron2013time}.
In addition, publicly available data from the Spanish National Statistics Institute are often not accessible at the level of disaggregation required for our analysis (i.e., by sex, age-group, year, and province). This limited the inclusion of additional potential risk factors, such as socioeconomic level, proportion of migrant population, antidepressant prescription rates, or access to mental health services.
Finally, although some recent studies have attempted to explore the impact of the COVID-19 pandemic on suicide mortality or suicide behavior \citep{dube2021suicide,fouillet2023reactive,lantos2024impact,ramalle2023trends,DELATORRELUQUE202429}, we find that it is still necessary to enlarge the study period to better understand the real impact of the pandemic. The span of our study from 2010 to 2022 makes it difficult to isolate and accurately assess these specific effects.


\section{Conclusion}\label{sec13}

This study highlights several key conclusions regarding suicide mortality in Spain from 2010 to 2022, emphasizing distinct patterns based on gender and geographical location. A notable and steady upward trend was observed in female suicide mortality, despite male suicide rates remaining consistently higher than female rates throughout the study period. The increase was particularly pronounced among younger age groups of females.

We also found that socioeconomic and contextual factors play a significant role in suicide rates. For males, suicide rates showed a statistically significant positive association with the proportion of the population living in rural areas, with a 10 percentage-point increase in rural population associated with an estimated 5.4\% rise in male suicide mortality. For females, the annual unemployment rate was a significant factor, with a one percentage-point increase linked to a 2.4\% rise in female suicide mortality.
While regional unemployment rates were associated with suicide mortality, these rates primarily reflect broader economic conditions rather than individual employment status. Thus, the observed associations should be interpreted as indicative of the general socioeconomic context influencing suicide risk, rather than as direct causal effects of unemployment at the individual level.

Geographic disparities were also evident, with males exhibiting greater heterogeneity in rates across provinces, while females displayed a smoother spatial distribution. Despite these differences, two regions, Galicia in the northwest and southern Andalusia, were consistently identified as having suicide rates markedly above the national mean for both sexes. The study also concluded that age is a major determinant of suicide risk, with the age group component accounting for the largest proportion of total variability in all models, explaining 99\% of the variability for males and 96\% for females.

\pagebreak

\subsection*{List of abbreviations}
CrI: credible interval, DIC: Deviance Information Criterion, iCAR: intrinsic conditional autoregressive, ICD: International Classification of Diseases, INE: National Statistics Institute (Spain), INLA: integrated nested Laplace approximations, RSR: restricted spatial regression, RW1: first-order random walk, RW2: second-order random walk, WAIC: Watanabe-Akaike Information Criterion.

\section*{Declarations}

\subsection*{Ethics approval and consent to participate}
Not applicable.

\subsection*{Consent for publication}
Not applicable.

\subsection*{Availability of data and materials}
All analyses in this paper are fully reproducible. The data and R code for fitting the models with R-INLA, as well as for generating all tables and figures, are publicly available at:
\url{https://github.com/spatialstatisticsupna/Suicides_Spain}.

\subsection*{Competing interests}
The authors declare no competing interests.

\subsection*{Funding}
This work is supported by projects PID2024-155382OB-I00 (funded by MICIU/AEI/10.13039/501100011033 and FEDER, UE), PID2020-113125RB-I00/MCIN/AEI/10.13039/501100011033 (Spanish Ministry of Science and Innovation), and UNEDPAM/PI/PR24/01A (Centro Asociado UNED - Pamplona).

\subsection*{Authors' contributions}
All authors contributed equally to this work.

\subsection*{Acknowledgements}
Not applicable.

\bigskip
\bibliographystyle{apalike}
\bibliography{refs}   


\counterwithin*{figure}{section}
\renewcommand{\thefigure}{S\arabic{figure}}
\renewcommand{\thetable}{S\arabic{table}}
\setcounter{figure}{0}

\clearpage
\section*{Supplementary material}
\addcontentsline{toc}{section}{Supplementary material}

\medskip
This supplementary material presents additional figures supporting the main analysis.
\bigskip

\begin{figure}[!ht]
    \begin{center}
    \includegraphics[width=\textwidth]{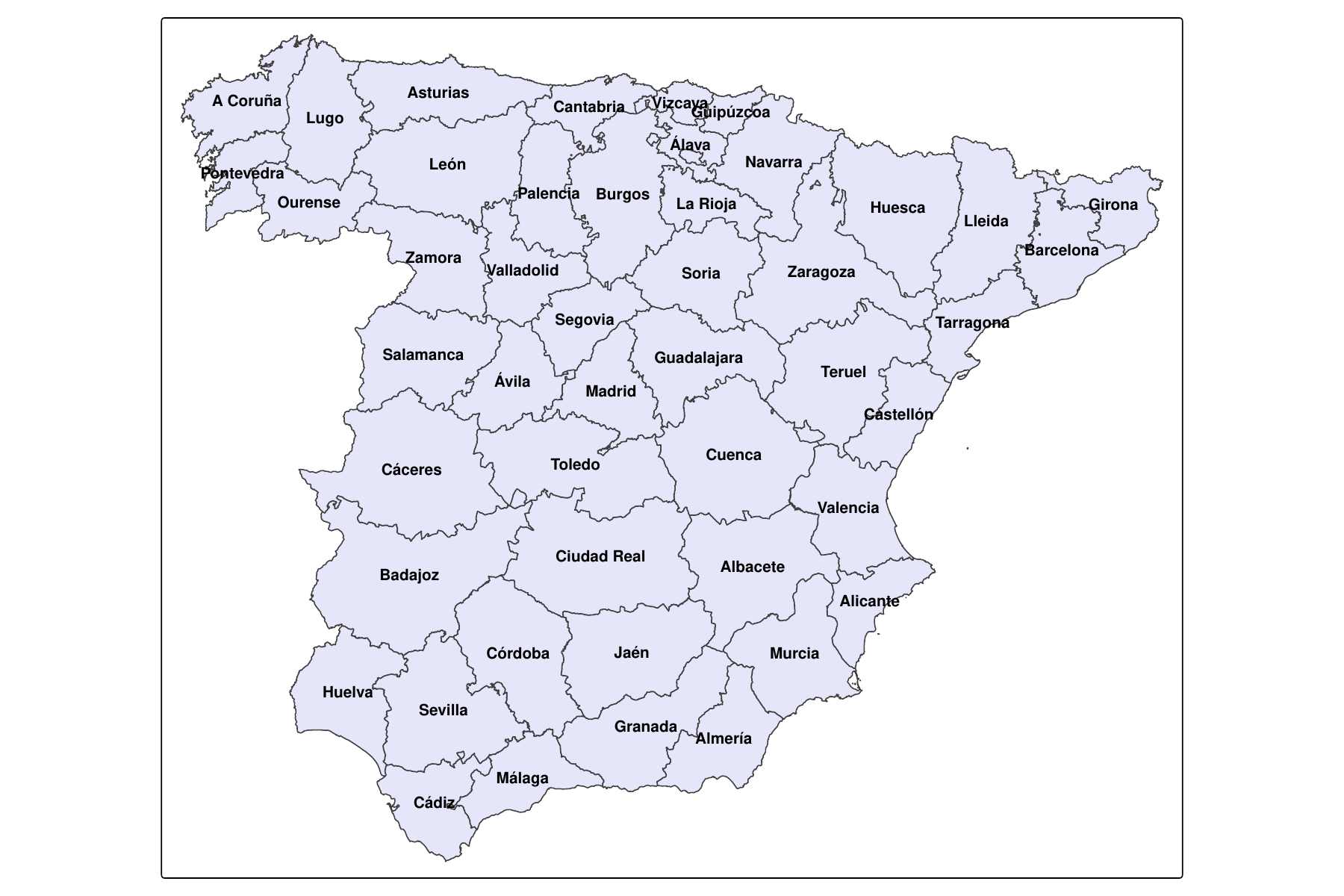}
    \end{center}
    \vspace{-0.2cm}
    \caption{Map with the administrative division of Spain by provinces.}
    \label{fig:Spain_provinces}
\end{figure}

\begin{figure}[!ht]
    \begin{center}
    \includegraphics[width=\textwidth]{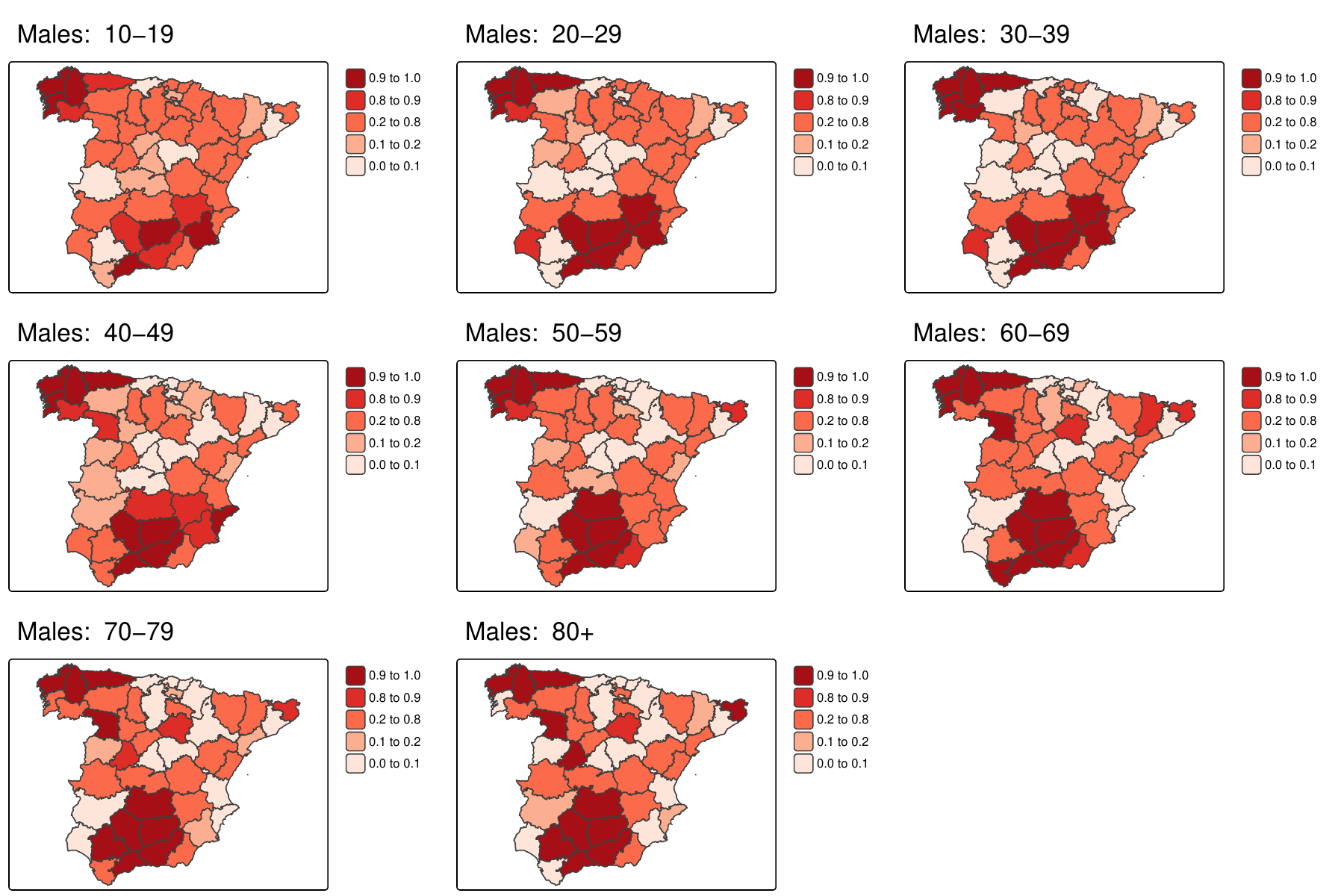}
    \end{center}
    \vspace{-0.2cm}
    \caption{Maps of posterior exceedance probabilities $Pr(\lambda_{as}>\hat{\alpha}+\hat{\phi}_a | {\bf O})$ for male population.}
    \label{fig:ExceedanceProbs_AgeSpace_Males}
\end{figure}

\begin{figure}[!ht]
    \begin{center}
    \includegraphics[width=\textwidth]{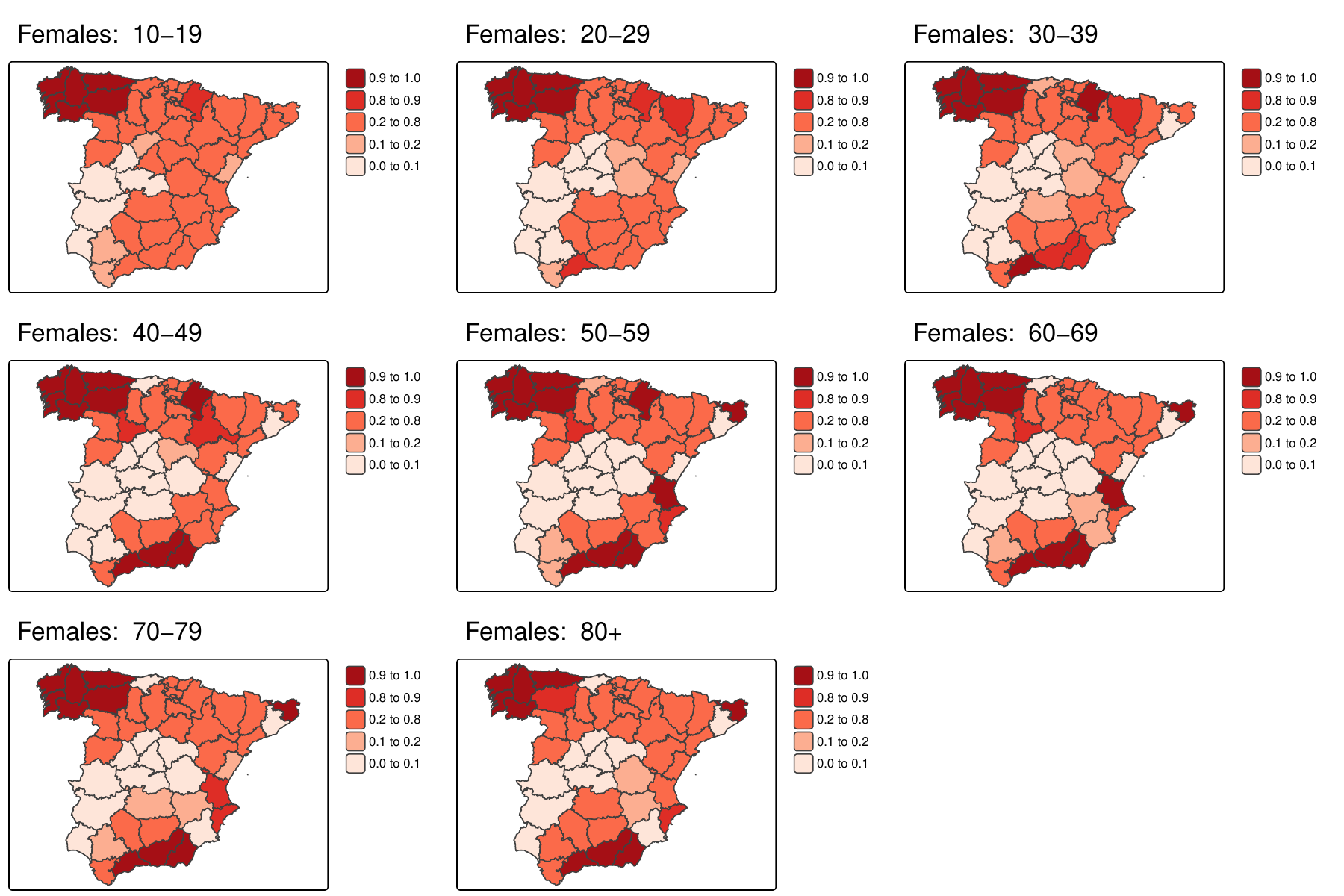}
    \end{center}
    \vspace{-0.2cm}
    \caption{Maps of posterior exceedance probabilities $Pr(\lambda_{as}>\hat{\alpha}+\hat{\phi}_a | {\bf O})$ for female population.}
    \label{fig:ExceedanceProbs_AgeSpace_Females}
\end{figure}

\end{document}